\documentclass[letter]{aa}         

\usepackage{natbib}
\usepackage{graphicx}
\usepackage{txfonts}
\usepackage{hyperref}
\usepackage{color}
\usepackage{upquote} 
\usepackage[normalem]{ulem}
\usepackage{xspace}

\def\Gaia{\textit{Gaia}\xspace}
\def\gmag{\ensuremath{G}\xspace}
\def\gbp{\ensuremath{G_{\rm BP}}\xspace}
\def\grp{\ensuremath{G_{\rm RP}}\xspace}

\newcommand{\mass}[1]{\ensuremath{#1\, \mathrm{M}_\odot}}

\title{A new method to identify subclasses among AGB stars using \Gaia and 2MASS photometry}
  \titlerunning{Classification of LPVs using \Gaia and 2MASS}
  \authorrunning{Lebzelter et al.}

\author{T. Lebzelter\inst{\ref{inst_vie}}\fnmsep\thanks{Corresponding author: T. Lebzelter
(\href{mailto:thomas.lebzelter@univie.ac.at}{\tt thomas.lebzelter@univie.ac.at})},
          N.~Mowlavi\inst{\ref{inst_gen}},
          P.~Marigo\inst{\ref{inst_pdv}},
          G.~Pastorelli\inst{\ref{inst_pdv}},          M.~Trabucchi\inst{\ref{inst_pdv}},
          P.R.~Wood\inst{\ref{inst_mso}},
          I.~Lecoeur-Ta\"{i}bi\inst{\ref{inst_gen}}
}

 \institute{University of Vienna, Department of Astrophysics, Tuerkenschanz\-strasse 17, A1180 Vienna, Austria\label{inst_vie}
 \and
 Department of Astronomy, University of Geneva, Ch. des Maillettes 51, CH-1290 Versoix, Switzerland\label{inst_gen}
 \and 
 Dipartimento di Fisica e Astronomia Galileo Galilei Università di Padova, Vicolo dell’Osservatorio 3, I-35122 Padova, Italy\label{inst_pdv}
  \and 
 Research School of Astronomy and Astrophysics, Australian National University, Canberra, ACT 2611, Australia\label{inst_mso}
       }

\date{June 2018}

\abstract{}{
    We explore the wealth of high quality photometric data provided by data release 2 of the \Gaia mission for long period variables (LPVs) in the Large Magellanic Cloud.
    Our goal is to identify stars of various types and masses along the Asymptotic Giant Branch.
}{
    For this endeavour, we developed a new multi-band approach combining Wesenheit functions $W_{RP,BP-RP}$ and $W_{K_s,J-K_s}$ in the \Gaia BP, RP and 2MASS J, K$_\mathrm{s}$ spectral ranges, respectively, and use a new diagram ($W_{RP,BP-RP}-W_{K_s,J-K_s}$) versus $K_s$ to distinguish between different kinds of stars in our sample of LPVs.
    We used stellar population synthesis models to validate our approach.
}{
    We demonstrate the ability of the new diagram to discriminate between O-rich and C-rich objects, and to identify low-mass, intermediate-mass and massive O-rich red giants, as well as extreme C-rich stars.
    Stellar evolution and population synthesis models guide the interpretation of the results, highlighting the diagnostic power of the new tool to discriminate between stellar initial masses, chemical properties and evolutionary stages.
}
{}

\begin{document}

\maketitle

\section{Introduction}
Asymptotic Giant Branch (AGB) stars represent a decisive part within the evolution of low and intermediate mass stars.
These high luminosity, cool objects contribute significantly to the metal content of their host galaxy by various nucleosynthesis processes in their interior, deep mixing events, and strong mass loss.
The first two points modify the surface composition -- even changing the chemistry from O-rich to C-rich -- which affects the mass loss.
All these contributions depend on the stellar parameters, in particular the stars' masses.

To study the role of AGB stars for the evolution of galaxies, it is important to reliably identify and characterize (mass, chemistry) them within a stellar population.
AGB stars show a typical variability pattern with long periods of several ten to several hundred days and moderate to large light amplitudes in particular in the visual range. 
These variables form the class of the long period variables (LPVs) which also includes other cool, evolved stars like red supergiants or stars at the tip of the Red Giant Branch (RGB).
This kind of variability is easily detectable even in other galaxies.
Thus, LPVs define an excellent sample of AGB stars.

Major advances in the understanding of AGB stars were often related to large surveys in various photometric bands.
Monitoring surveys like MACHO \citep{1999IAUS..191..151W} and OGLE \citep[e.g.][]{2009AcA....59..239S} and others delivered excellent samples of AGB stars for various galaxies and the Galactic Bulge. 
With the advent of the \Gaia mission \citep{2016A&A...595A...1G}, a new era is starting.
The all-sky nature of the \Gaia survey, the high spatial resolution of $\sim$0.4~arcsec, the provision of parallaxes, the availability of three photometric bands providing wide \gmag, blue \gbp and red \grp magnitudes for more than 1.5~billion stars, from very bright (few magnitudes in \gmag) to as faint as $\gmag \simeq 20.7$~mag, and the provision of time series in all three bands for all the stars, not mentioning the expected provision of radial velocity and astrophysical parameters for stars brighter than $\gmag \simeq 16$~mag, are some of the unique advantages of the survey.

In \Gaia data release 2 \citep[DR2,][]{Brown_etal18},
the first \Gaia catalog of LPVs with \gmag-band variability amplitudes larger than 0.2~mag (measured by the 5--95\% quantile range of the \gmag time series) has been published \citep{2018LPV_DR2}.
It contains 151\,761 candidates.
As a first step in the analysis of this unique dataset, we study in this Letter the \Gaia DR2 LPVs in the Large Magellanic Cloud (\object{LMC}).

The above mentioned task of identification and distinction of various groups of stars among the LPVs is challenging because the upper giant branch represents a mixture of objects of various masses and evolutionary stages.
In this letter we present a new method for an efficient identification of stars of different mass and chemistry based on \Gaia and 2MASS photometry. 

\section{LPV characterization using \Gaia and 2MASS photometry} 
\label{Sect:LPVcharacterization}

\begin{figure}
\centering
\includegraphics[width=\hsize]{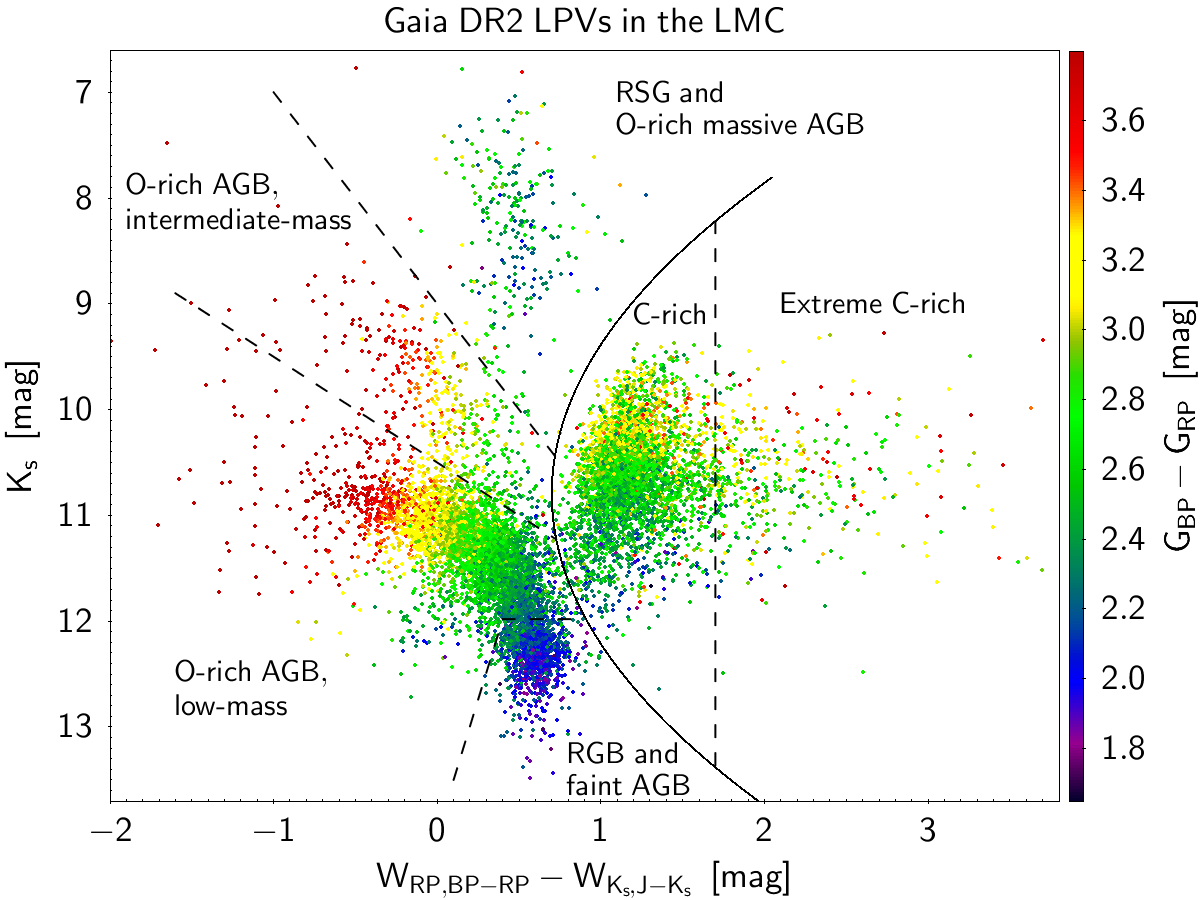}
\caption{$(W_{RP,BP-RP}-W_{K_s,J-K_s})$ versus $K_s$ diagram of \Gaia DR2 LPVs in the LMC.
         The markers are coloured with $\gbp-\grp$ according to the colour-scale shown on the right of the figure.
         The solid line delineates O-rich (left of the line) and C-rich (right of the line) stars, and dashed lines delineate sub-groups as indicated in the figure.
         The definition of the borderlines are given in Appendix~\ref{Sect:Wesenheit_Gaia}.
        }
\label{Fig:LMC_WRPminusWK_versus_K_withBPmRP}
\end{figure}

The most efficient way to distinguish various types of stars in large samples like a galaxy is via photometry. 
Frequently used in this context is the distinction between O-rich and C-rich stars by using near-infrared photometry ($J-K_s$) from the 2MASS all sky catalogue \citep{2006AJ....131.1163S}.
Cross-matches within a search radius of 1$\arcsec$ between \Gaia and 2MASS sky positions could be obtained for all the 11\,022 \Gaia LMC LPVs studied here (see below for the selection of LMC candidates), except for only seven candidates. 

The $K_s$-band is a quite good representative of the bolometric magnitude for red giants.
To have a reddening free magnitude, we utilized the near infrared Wesenheit function $W_{K,J-K} = K_s - 0.686\,\cdot\,(J-K)$ \citep[in the following $W_K$, see][]{2005AcA....55..331S}. 
Several authors have already used $W_K$, e.g.~for the construction of  period-luminosity diagrams of LPVs \citep[e.g.][]{2017ApJ...847..139T}.

The intrinsic colour spread of red giant stars being relatively small in the infrared -- except for large values of circumstellar reddening when gas and dust expelled from the star surround the object 
--, the Wesenheit function $W_K$ can adequately correct for interstellar reddening using $J-K_s$, since interstellar reddening impacts both the $K_s$ magnitude and the $J-K_s$ color.

In the visual range, the colour of red giants is very sensitive to the surface temperature, and there is a large intrinsic visual colour spread in a population of red giants.
This is in particular the case for the \Gaia $\gbp-\grp$ colour that we use in this study, since
$\gbp-\grp$ is strongly affected by the temperature sensitivity of molecules dominating these wavelength ranges.
The molecules present in the atmosphere depend on the C/O ratio.
Therefore, a degeneracy exists between interstellar reddening, circumstellar reddening, chemistry, and temperature differences at a given luminosity.
By using Wesenheit functions, we can eliminate the interstellar reddening component (and much of the circumstellar reddening component) from this degeneracy.

We now define the Wesenheit function $W_{RP}$, a reddening-free combination of \gbp and \grp magnitudes.
The factor multiplying $(\gbp-\grp)$ in $W_{RP}$ can be computed using the interstellar extinction data provided by \citet{1998ApJ...500..525S}, adapted for the BP and RP filter curves.
As a result\footnote{We use the median values of \gbp and \grp provided in \texttt{gaiadr2.vari\_time\_series\_statistics} at \url{https://gea.esac.esa.int/archive/}}, we get 
\begin{equation}
   W_{RP} = \grp - 1.3\,\cdot\,(\gbp-\grp).
\end{equation}
In LPVs, 
$W_{RP}$ combines temperature, chemistry (O- or C-rich), reddening and brightness information. 
This is not the case for $W_K$, the colour $J-K_s$ being much less sensitive to the surface temperatures and chemistry of LPVs than $\gbp-\grp$ is. 
We will come back to this point below.

\begin{figure}
\centering
\includegraphics[width=\hsize]{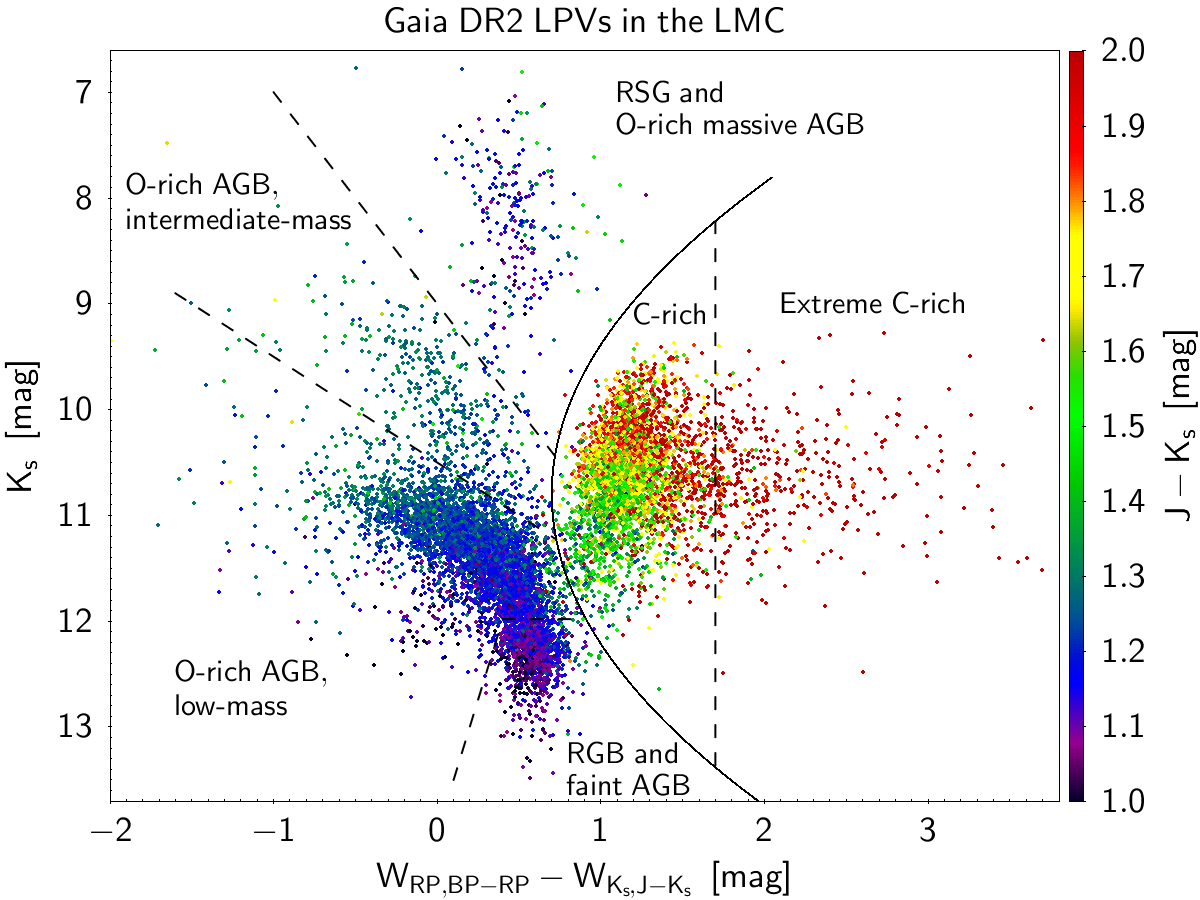}
\caption{Same as Fig.~\ref{Fig:LMC_WRPminusWK_versus_K_withBPmRP}, but with the markers coloured with $J-K_s$ according to the colour-scale shown on the right of the figure. 
        }
\label{Fig:LMC_WRPminusWK_versus_K_withJmK}
\end{figure}

In a next step we attempted to combine the three quantities $K_s$, $W_{RP}$ and $W_K$ in a 2-dimensional diagram by combining $W_{RP}$ and $W_K$.
In Fig.\,\ref{Fig:LMC_WRPminusWK_versus_K_withBPmRP} we plot $W_{RP}-W_K$ on the x-axis and  $K_s$ on the y-axis for all \Gaia DR2 LPVs identified in the LMC.
The selection of members is based on
sky position ($ 50^{\circ} < \alpha < 105^{\circ}$ and $-77^{\circ} < \delta < -61^{\circ}$),
proper motion ($ 1.2 < \mu_\alpha^* < 2.5$ and $-0.8 < \mu_\delta < 1.5$, in mas/yr)
and parallaxes ($\varpi < 0.5$~mas), 
which results in 11\,022 LMC candidates.
As described in more detail in Appendix \ref{Sect:Wesenheit_Gaia}, it is important to understand that  Fig.\,\ref{Fig:LMC_WRPminusWK_versus_K_withBPmRP} is not a CMD in the classical sense since the bluest objects are located in the middle of the diagram, around $W_{RP}-W_K=0.8$~mag for the LMC, and the objects are getting redder towards both sides of the diagram.
This is illustrated, for the left side of the diagram, in Fig.\,\ref{Fig:LMC_WRPminusWK_versus_K_withBPmRP} where the colour-coding of the points shows that the stars have an increasing $\gbp-\grp$ colour (they become redder) when $W_{RP}-W_K$ decreases to values below $W_{RP}-W_K=0.8$~mag.
Figure~\ref{Fig:LMC_WRPminusWK_versus_K_withJmK}, on the other hand, which plots the same diagram but with a colour-code related to $J-K_s$, shows that, on the right side of the diagram, stars have an increasing $J-K_s$ colour (they also become redder) when $W_{RP}-W_K$ increases to larger values above $W_{RP}-W_K = 0.8$~mag.

\section{Applications of the index $W_{RP}-W_K$}
The $(W_{RP}-W_K)$ versus $K_s$ diagram turns out to be a very powerful tool to discriminate between various types of AGB stars.
This mainly results from two facts.
First is the fact that C-rich stars are distinguished from O-rich stars from their redder $J-K_s$ colours \citep[e.g.][]{2003A&A...403..225M}%
\footnote{The \Gaia pipeline did not discriminate between C and M type stars in DR2 \citep{2018LPV_DR2}}.
In the $(W_{RP}-W_K)$ versus $K_s$  diagram, this implies that C-rich stars will be located on the right side of the diagram, at $W_{RP}-W_K \gtrsim 0.8$~mag for the LMC (as confirmed from the location of these stars in the 2MASS colour-magnitude diagram, see below), while O-rich stars are expected to be on the left at $W_{RP}-W_K \lesssim 0.8$~mag.
This nice discrimination between C-rich and O-rich chemistry in the diagram is a consequence of distinct molecular absorption features present in the spectra of these two types of stars (see Appendix~\ref{Sect:Wesenheit_Gaia} for more details).
The second fact that allows to discriminate various types of AGB stars in the diagram is related to their specific evolution properties as a function of their initial stellar mass, $M_i$.
Schematically, AGB stars evolve to C-rich stars only for initial stellar masses above a given threshold\footnote{The mass limit for the formation of C-rich stars significantly depends on model details, in particular the third dredge-up efficiency, and varies from author to author. Here we report the results from our ongoing calibration of the TP-AGB phase.} \citep[$M_i \gtrsim$ \mass{1.45} at Z=0.006, see e.g.][]{Marigo_etal_2013}.
Intermediate-mass AGB stars (\mass{M_i \gtrsim 3.0-3.5} at Z=0.006, depending on model details), on the other hand, do generally not become C-stars due to mass loss and/or hot-bottom burning (HBB).
These different types of AGB stars are hence expected to populate distinct regions in the $(W_{RP}-W_K)$ versus $K_s$ diagram.

\begin{figure}
\centering
\includegraphics[width=\hsize]{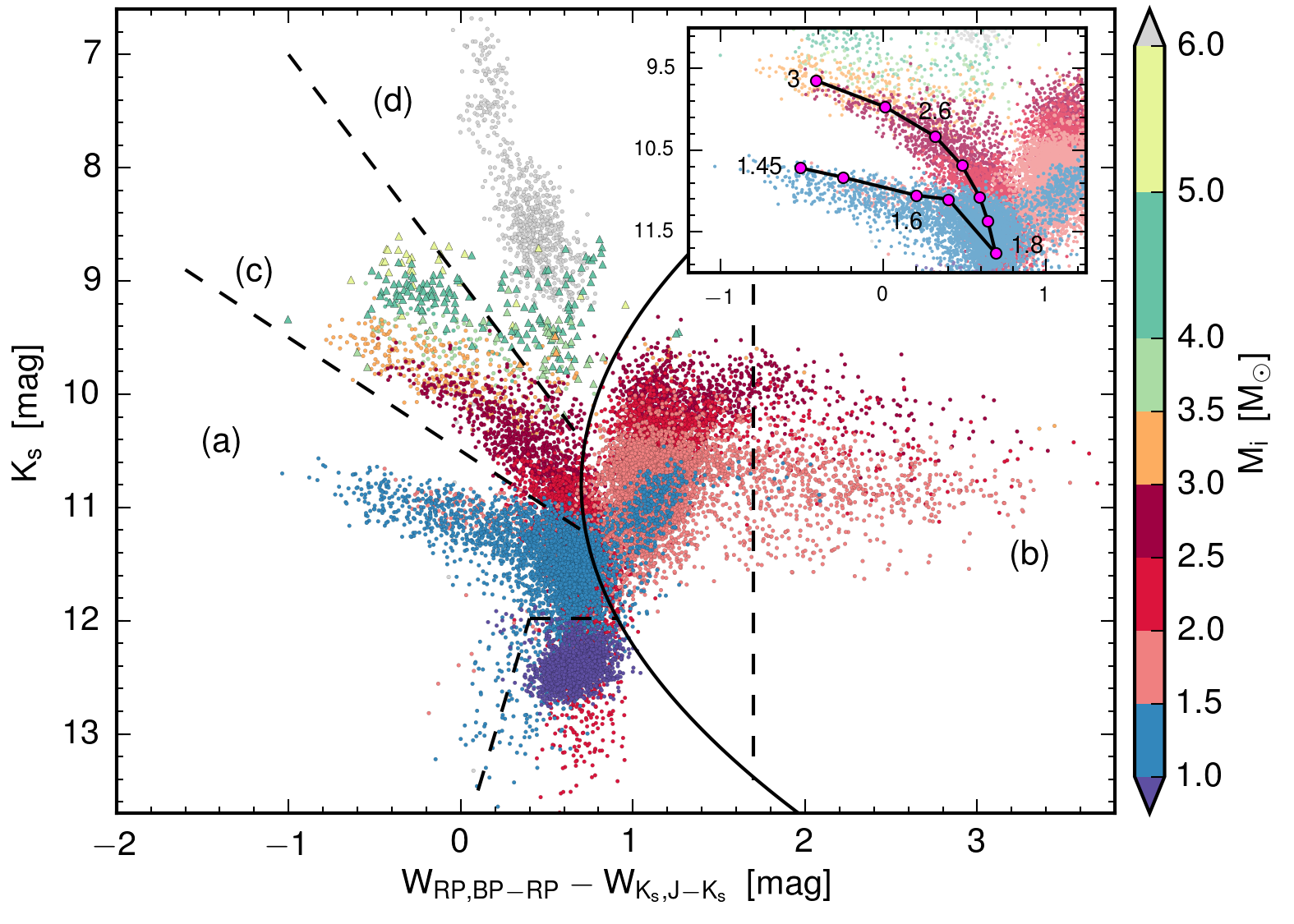}
\caption{Synthetic stellar populations for the LMC, colour-coded according to the initial stellar mass. The simulation is obtained with the \texttt{TRILEGAL} code \citep{Girardi_etal_2005} and stellar isochrones that include a detailed description of the AGB phase \citep{Marigo_etal_2013, Marigo_etal_2017}. Suitable selection criteria for LPVs are applied, based on new pulsation models \citep{2017ApJ...847..139T}. Letters a, b, c, and d identify the main branches of evolved stellar populations. In the inset the curve connects the stages just before stars turn to C-stars at varying initial mass in the range from 1.45 $M_{\odot}$ to 3 $M_\odot$ (few values in $M_{\odot}$ are indicated; the initial metallicity is  Z=0.006). Triangles are used to mark AGB stars with HBB.
See text and Appendix~\ref{appendix:models} for more explanation.
The dashed and solid lines correspond to the same empirical boundaries as in Fig.~\ref{Fig:LMC_WRPminusWK_versus_K_withBPmRP}.
}
\label{Fig:model_LMC_WRP_WJK}
\end{figure}

\begin{figure}
\centering
\includegraphics[width=\hsize]{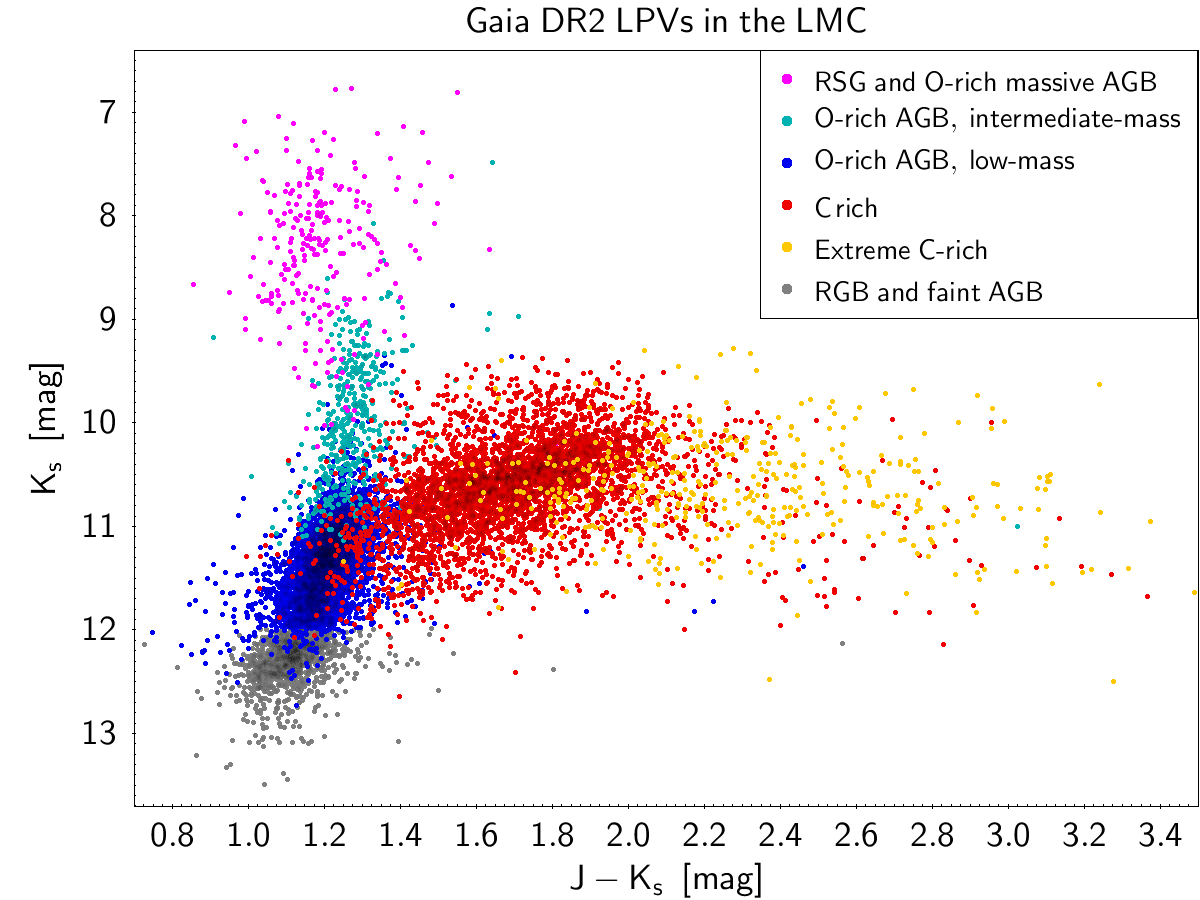}
\caption{2MASS colour-magnitude diagram $(J-K_s)$ versus $K_s$ of \Gaia DR2 LPVs in the LMC.
         The colour of the symbols are related to the groups of stars identified in Fig.~\ref{Fig:LMC_WRPminusWK_versus_K_withBPmRP} according to the colour-coding written in the inset.
         The axis ranges have been limited for better visibility.
        }
\label{Fig:LMC_CM_2mass}
\end{figure}

A total of six groups are identified in the $(W_{RP}-W_K)$ versus $K_s$ diagram, as shown in Figs.~\ref{Fig:LMC_WRPminusWK_versus_K_withBPmRP} and \ref{Fig:LMC_WRPminusWK_versus_K_withJmK}.
Of these groups, four consist of O-rich stars, located at the left of the solid line on the diagram, while the remaining two groups on the right of the solid line contain the C-rich stars.
The separation of C-rich stars in two distinct groups reflects the drop in
concentration of stars beyond $W_{RP}-W_K\sim1.7$~mag (see discussion below).
Throughout the remainder of this letter, we refer to
these objects as \textit{C-rich} and \textit{Extreme C-rich stars}, respectively.
The term ``extreme'' stars was originally introduced by \citet{Blum_etal_2006} 
in their Spitzer SAGE survey of the  LMC to identify a group of very red objects in near-mid infrared CMDs, interpreted
mostly as dust-enshrouded C-rich stars. 
It should be clear that our criterion for extreme C-rich stars is not the same, but spots out a sub-sample of the extreme objects in the original definition.

The identification of subgroups among the O-rich M-type stars follows the structures suggested in Figs.~\ref{Fig:LMC_WRPminusWK_versus_K_withBPmRP} and \ref{Fig:LMC_WRPminusWK_versus_K_withJmK}.
The tip of the RGB is located close to $K_s=$12 in the LMC \citep{cioni_et_al_2000}.
We therefore assume that the stars in the group 'RGB and faint AGB' are primarily tip-RGB stars or stars on the early AGB.
For the remaining, brighter, O-rich stars, three subgroups are distinguishable in the diagram, as delineated by the dashed lines on the left of the solid line in the two figures.
Comparison with model predictions, discussed below, reveals that they are associated with low-mass, intermediate-mass, and massive O-rich AGB stars, respectively.

In order to interpret the observed $(W_{RP}-W_K)$ versus $K_s$ diagram, we compute stellar population synthesis models of the LMC that take into account the stellar and galactic properties of the Cloud (see Appendix~\ref{appendix:models} for details).
The result of the simulation is shown in Fig.~\ref{Fig:model_LMC_WRP_WJK}, with the points colour-coded according to the initial mass of the simulated stars. 
The simulated diagram (Fig.~\ref{Fig:model_LMC_WRP_WJK}) shows a similar structure as the observed one (Fig.~\ref{Fig:LMC_WRPminusWK_versus_K_withBPmRP}). 
In particular the  segregation of stars in the various branches is well recovered by models.
The colour-coding clearly indicates that the various groups visible in Fig.~\ref{Fig:LMC_WRPminusWK_versus_K_withBPmRP} can be directly related to various mass ranges (for further comparisons, see  Fig.~\ref{Fig:LMC_WRPminusWK_versus_K_withTracks} of the Appendix that nicely associates the branches visible in the diagram to stellar evolution tracks of various masses).

The comparison of the observed (Figs.~\ref{Fig:LMC_WRPminusWK_versus_K_withBPmRP} and \ref{Fig:LMC_WRPminusWK_versus_K_withJmK}) and simulated (Fig.~\ref{Fig:model_LMC_WRP_WJK}) diagrams confirms that\footnote{The mass limits given in this Letter for the different groups are derived from our LMC simulations}:
\begin{itemize}
\item the group 'O-rich AGB, low-mass' in Fig.~\ref{Fig:LMC_WRPminusWK_versus_K_withBPmRP} mainly contains low-mass O-rich stars during the early-AGB and thermally pulsing AGB (TP-AGB) phases, with initial masses from $\sim$\mass{0.9} to $\sim$\mass{1.4} (branch \textit{(a)} in Fig.~\ref{Fig:model_LMC_WRP_WJK}).
These stars do not become carbon stars (in these calculations) and extend to the negative $W_{RP}-W_{K}$ values which they reach at the end of their AGB evolution. A small number of stars in the mass range $\sim$\mass{1.4} to $\sim$\mass{1.8} lie on the bright side of branch \textit{(a)} when they are in the evolutionary stage just prior to becoming carbon stars;
\vskip 1mm
\item the group 'C-rich' in Fig.~\ref{Fig:LMC_WRPminusWK_versus_K_withBPmRP} contains carbon stars (C/O>1, branch \textit{(b)}) with distinct $J-K_s$ colours, see Fig.~\ref{Fig:model_LMC_WRP_WJK}. In our simulation of the LMC, they are characterized by initial masses from $\sim$\mass{1.4} to $\sim$\mass{3.2};
\vskip 1mm
\item the group 'Extreme C-rich' (C-stars with very large $J-K_s$ colours) in Fig.~\ref{Fig:LMC_WRPminusWK_versus_K_withBPmRP} is an extension of the group of C-rich stars in branch \textit{(b)}. In the simulation they are characterized by strong mass loss driven by radiation pressure on carbonaceous dust grains. The relatively low concentration of such objects in the observed diagram is mainly due to the short timescales towards the termination of the AGB phase.

\vskip 1mm
\item the group 'O-rich AGB, intermediate-mass' in Fig.~\ref{Fig:LMC_WRPminusWK_versus_K_withBPmRP} (branch \textit{(c)}) actually contains two kinds of stars.
On the one hand, the faint edge of the branch is populated by \mass{M_i\gtrsim 2} to $\sim$\mass{3.2} stars that are still O-rich (stars between $\sim$\mass{1.8} and $\sim$\mass{2} do not move into branch \textit{(c)} before becoming C-rich).
They will move to branch \textit{(b)} when they turn into C-stars later along the TP-AGB phase.
The bright edge of the branch, on the other hand, hosts intermediate-mass stars that never become C-stars, with $M_i$ from $\sim$\mass{3.2} to $\sim$\mass{6}. 
These include  AGB stars that experience HBB (see Fig.\ref{Fig:model_LMC_WRP_WJK}).
We note that during the very last, extremely short-lived,  evolutionary stages characterized by very high mass loss rates and efficient dust production, these O-rich intermediate-mass AGB stars can move into the right part of the diagram slightly contaminating the group of Extreme C-rich stars. 
For further discussion we refer to Appendix \ref{appendix:models};
\vskip 1mm
\item the paucity of stars between branch \textit{(a)} (low-mass O-rich) and branch \textit{(b)} (intermediate-mass O-rich) reflects the evolution prior to the AGB, besides the properties of the third dredge-up. 
To help the analysis, in the inset of Fig.~\ref{Fig:model_LMC_WRP_WJK}
we show the predicted location of stars with initial masses in the range $\sim$\mass{1.4} to $\sim$\mass{3.0} at the stage just before becoming C-stars at the next thermal pulse. As is evident, the curve contours the gap between branches \textit{(a)} and \textit{(c)}.
The fainter section of the curve is drawn by low-mass stars that
developed an electron degenerate He-core after the main sequence ($\sim$\mass{1.4} to $\sim$\mass{1.8}), while 
the brighter section is traced by intermediate-mass stars that avoided core degeneracy ($\sim$\mass{1.8} to $\sim$\mass{3}). 
Stars at the transition between the low- and intermediate-mass classes, i.e. with \mass{M_i \simeq 1.8} (see inset of Fig.~\ref{Fig:model_LMC_WRP_WJK}), ignite He at the smallest core mass of all, and therefore also have the lowest luminosity. With the current TP-AGB models, they also have the faintest subsequent transition luminosity to the C-star domain;
\vskip 1mm
\item finally, the group 'RSG and O-rich massive AGB' in Fig.~\ref{Fig:LMC_WRPminusWK_versus_K_withBPmRP} is mainly composed of red supergiant stars (\mass{M_i \gtrsim 8}) burning He in their core (branch \textit{(d)} in Fig.~\ref{Fig:model_LMC_WRP_WJK}).
Our models suggest that a small fraction of them, distributed around the faint side of the group, also consist of \mass{M_i \simeq 6} to $\sim$\mass{8} stars that include super-AGB stars \citep{Siess_2010}, 
as well as massive AGB stars with $M_i$ from $\sim$\mass{5} to $\sim$\mass{6}.
\end{itemize}

The $(W_{RP}-W_K)$ versus $K_s$ diagram is thus a powerful tool to identify the different types of AGB stars.
Compared to the classical 2MASS colour-magnitude diagram, shown in Fig.~\ref{Fig:LMC_CM_2mass} with the various groups highlighted in colour, it provides the power to discriminate AGB stars according to their initial masses in addition to their chemistry.

\begin{figure}
\centering
\includegraphics[width=\hsize]{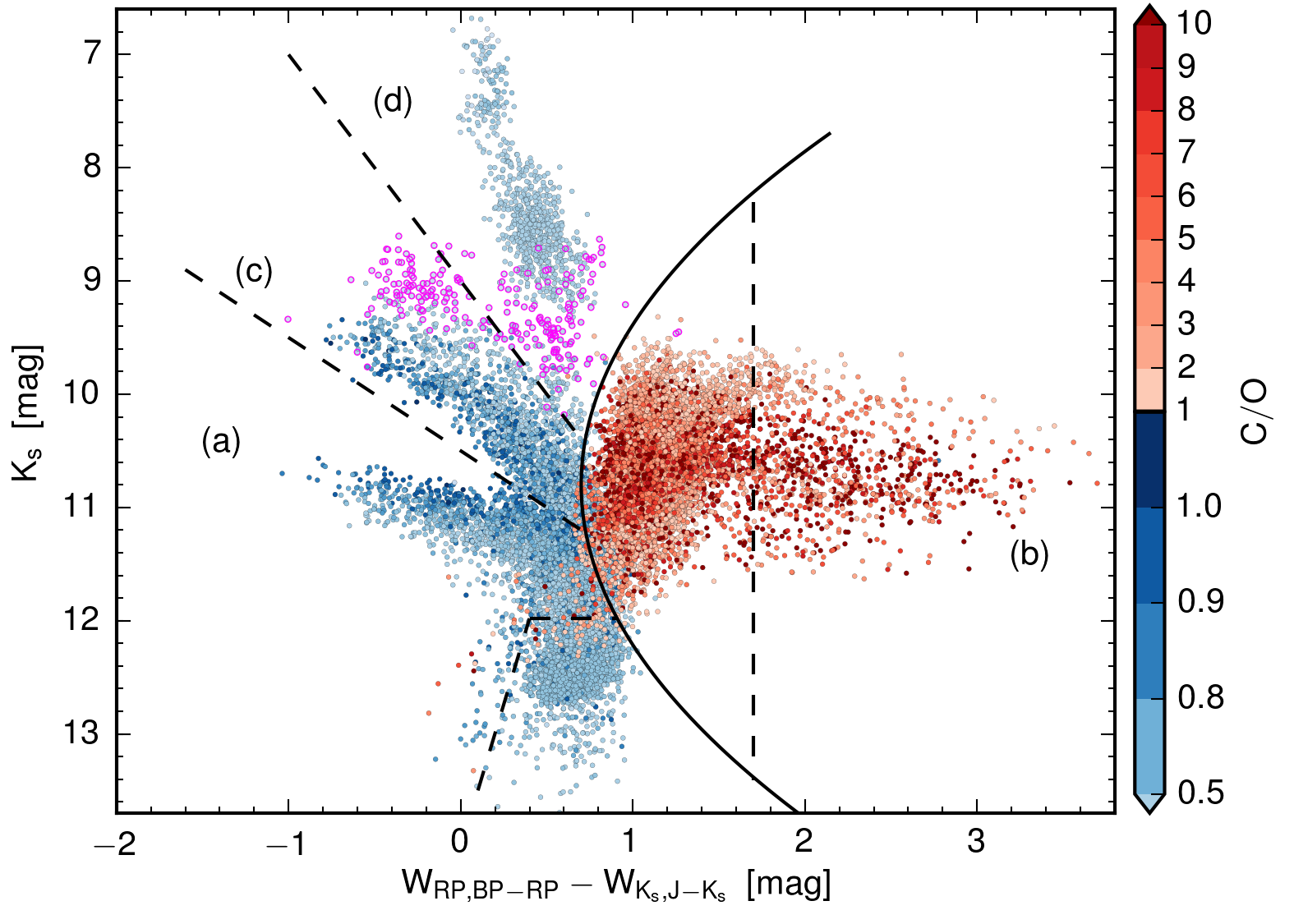}
\caption{Same as Fig.~\ref{Fig:model_LMC_WRP_WJK}, colour-coded according to the C/O ratio. AGB stars with HBB are highlighted with magenta circles.}
\label{Fig:model_CO_LMC_WRP_WJK}
\end{figure}

Finally, we note an interesting feature in the observed $(W_{RP}-W_K)$ versus $K_s$ diagram of the LMC.
There seems to be a real gap between the O-rich and C-rich clumps in Figs.~\ref{Fig:LMC_WRPminusWK_versus_K_withBPmRP} and \ref{Fig:LMC_WRPminusWK_versus_K_withJmK}.
This results from the fast transition between O-rich and C-rich regimes as soon as the C/O ratio reaches unity, as a consequence of the abrupt change in the molecular abundance pattern in the atmospheres of these stars \citep[e.g.,][]{2009A&A...508.1539M}. This is clearly visualized
in Fig.~\ref{Fig:model_CO_LMC_WRP_WJK} where we show the simulated $W_{RP}-W_K$ versus $K_s$ of the LMC, colour-coded according to the predicted C/O ratio. Note that the upper part of branch \textit{a} and the lower part of branch \textit{c} contain O-rich stars with values of the C/O close to one (darker blue points in Fig.~\ref{Fig:model_CO_LMC_WRP_WJK}). These should correspond to stars of spectral type S. HBB stars on the bright part of the diagram are characterized by low C/O ratios, N-enhancement, and low $^{12}$C/$^{13}$C.

%
\section{Conclusions}
\label{Sect:conclusions}

We demonstrated that the new tool for the investigation of stellar populations of AGB-stars presented in this paper offers an excellent possibility to study the large \Gaia dataset of evolved stars. It allows to distinguish stars of different mass, chemistry and evolutionary stage very efficiently. 
The high diagnostic capability of the new diagram is corroborated with the help of state-of-the-art stellar evolution models.

We plan a more extensive exploration of this tool in the future. 
As a next step, we are going to use this tool for a similar analysis of other stellar populations and for an investigation on the pulsational behaviour of LPVs.


\begin{acknowledgements}
We thank the referee for her/his suggestions on sharpening the focus of our letter.
PM, GP, MT acknowledge the support from the ERC Consolidator Grant funding scheme ({\em project STARKEY}, G.A. n. 615604).
\end{acknowledgements}

\bibliographystyle{aa}
\bibliography{LPV_PLD}

\appendix

\section{A new diagram for distinguishing various groups of stars on the AGB}
\label{Sect:Wesenheit_Gaia}

\begin{table*}
\caption{Definition of the different groups identified in the $(W_{RP,BP-RP}-W_{K,J-K})$ versus $K_s$ diagram of \Gaia DR2 LPVs (cf. Figs.~\ref{Fig:LMC_WRPminusWK_versus_K_withBPmRP} and \ref{Fig:LMC_WRPminusWK_versus_K_withJmK}), valid for the LMC.
        }
\centering
\begin{tabular}{l l}
\hline\hline
Group candidates & Definition for the LMC\\
\hline
O-rich                   & $W_{RP}-W_K \leq 0.7+0.15*(K_s-10.8)^2$ \\
-- faint AGB and RGB     & O-rich and $K_s \geq 11.98$ and $K_s \geq 13.5-5.067*(W_{RP}-W_K-0.1)$ \\
-- low-mass AGB          & O-rich and $K_s > 10.5+(W_{RP}-W_K)$ and [$K_s < 11.98$ or $K_s < 13.5-5.067*(W_{RP}-W_K-0.1)$] \\
-- intermediate-mass AGB & O-rich and $K_s > 9+2*(W_{RP}-W_K)$ and $K_s <= 10.5+(W_{RP}-W_K)$\\
-- massive AGB and RSG   & O-rich and $Ks <= 9+2*(W_{RP}-W_K)$\\
\hline
C-rich                   & $W_{RP}-W_K > 0.7+0.15*(K_s-10.8)^2$ \\
-- non Extreme C-star    & C-rich and $W_{RP}-W_K <= 1.7$ \\
-- Extreme C-star        & C-rich and $W_{RP}-W_K > 1.7$ \\
\hline
\end{tabular}
\label{Tab:groupsDefinition}
\end{table*}

\begin{figure*}
\centering
\includegraphics[width=0.95\hsize]{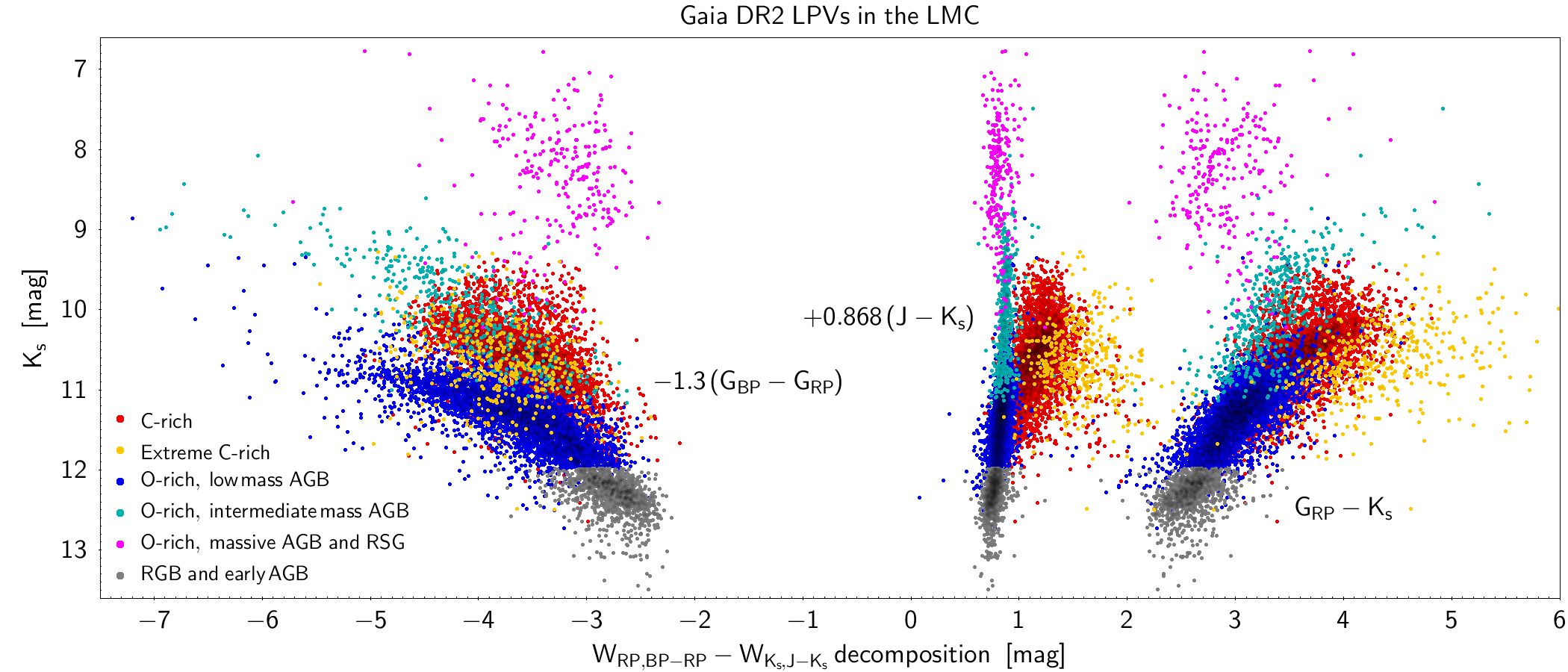}
\caption{The $W_{RP}-W_K$ index decomposed into its three contributing colours.
The points with negative values on the horizontal axis represent the term  $-1.3\;(\gbp-\grp)$, the points between 0 and $\approx$2 on the horizontal axis represent the term $+0.686 \; (J-K_s)$ while the points more positive than $\approx$2
represent the term $G_{RP} - K_s$.
         Different stellar groups are marked with different colours as indicated in the lower left corner of the figure.
         See text for a detailed description.
        }
\label{Fig:LMC_WRPminusWK_decomposition}
\end{figure*}

In Sect.~\ref{Sect:LPVcharacterization}, we presented a new diagram to study stars on the upper giant branch. 
The diagram uses a combination of two Wesenheit functions for the x-axis and the $K_s$-band brightness for the y-axis.
As shown in Fig.~\ref{Fig:LMC_WRPminusWK_versus_K_withBPmRP}, six distinct groups of red giants have been identified.
The definition of the borders of the groups in this diagram is given in Table~\ref{Tab:groupsDefinition}.
In this appendix, we elaborate more on the newly introduced $W_{RP}$ Wesenheit function and $W_{RP}-W_{K}$ index.

The index $W_{RP}-W_K$ actually combines three colours, and cannot be interpreted simply as a colour itself.
Instead, objects are getting redder both towards lower and higher values of that index with the bluest object being found, for the LMC, close to a value of 0.8.

To understand why this diagram can be used to disentangle various groups on the upper giant branch, we explored the three colour indices that enter $W_{RP}-W_K$ separately. 
Fig.~\ref{Fig:LMC_WRPminusWK_decomposition} shows the running of each of the three colours individually.
Note that $\gbp-\grp$ enters this plot with a negative sign, since it is subtracted in the combined formula. 
The sum of the three components as plotted here gives $W_{RP}-W_K$.

For the low mass O-rich stars (blue dots), the $W_{RP}-W_K$ value is mainly determined by $\gbp-\grp$ plus a contribution from $\grp-K_s$.
$J-K_s$ adds an almost constant value only.
In the BP ($\sim$330~--~680~nm) and RP ($\sim$640~--~1050~nm) wavelengths range, the spectra of O-rich stars are dominated by molecular bands from TiO, VO, ZrO, and VO. 
All these molecular bands increase in strengths with decreasing temperature.
The strong effect we see in this colour stems from the fact that the flux at the longward end ($\sim$1~$\mu$m) of the RP filter is only mildly affected by molecular absorption.
This is nicely illustrated in Fig.\,1 of \citet{2016MNRAS.457.3611A} showing model spectra in this wavelength range for various values of $T_{\rm eff}$.
Accordingly, the change in $W_{RP}-W_K$ from 0.5 to $-$1.5 seen in Fig.\,\ref{Fig:LMC_WRPminusWK_versus_K_withBPmRP} indicates a change towards \emph{lower} temperature.

A very similar situation is seen for the intermediate-mass O-rich stars (cyan points in Fig.~\ref{Fig:LMC_WRPminusWK_decomposition}), which is not surprising since the same molecules will affect the spectrum.
Massive AGB stars and supergiants (magenta points), on the other hand, show smaller changes in all three colours compared to the other groups. This is in agreement with red supergiants exhibiting on the average higher $T_{\rm eff}$ values  \citep{2016A&A...592A..16D} compared to low- and intermediate-mass AGB stars. 
As a consequence, this group is found at an almost constant value in $W_{RP}-W_K$ in Fig.\,\ref{Fig:LMC_WRPminusWK_versus_K_withBPmRP}

For C-stars (red points), the $W_{RP}-W_K$ value is almost identical to $J-K_s$. 
The other two colours cancel out each other since they have almost exactly the same value. 
$\gbp-\grp$ is dominated by CN and C$_2$ bands in this case, while the $K_s$ band is only mildly affected by C$_2$, CN, and CO \citep[compare Fig.\,3 in][]{2017A&A...601A.141G}.

\section{Synthetic Stellar Populations for the LMC: photometry and long-period variables}
\label{appendix:models}

\subsection{Stellar evolution tracks in the $W_{RP}-W_K$ versus $K_s$ diagram}

\begin{figure*}
\centering
\includegraphics[width=0.8\hsize]{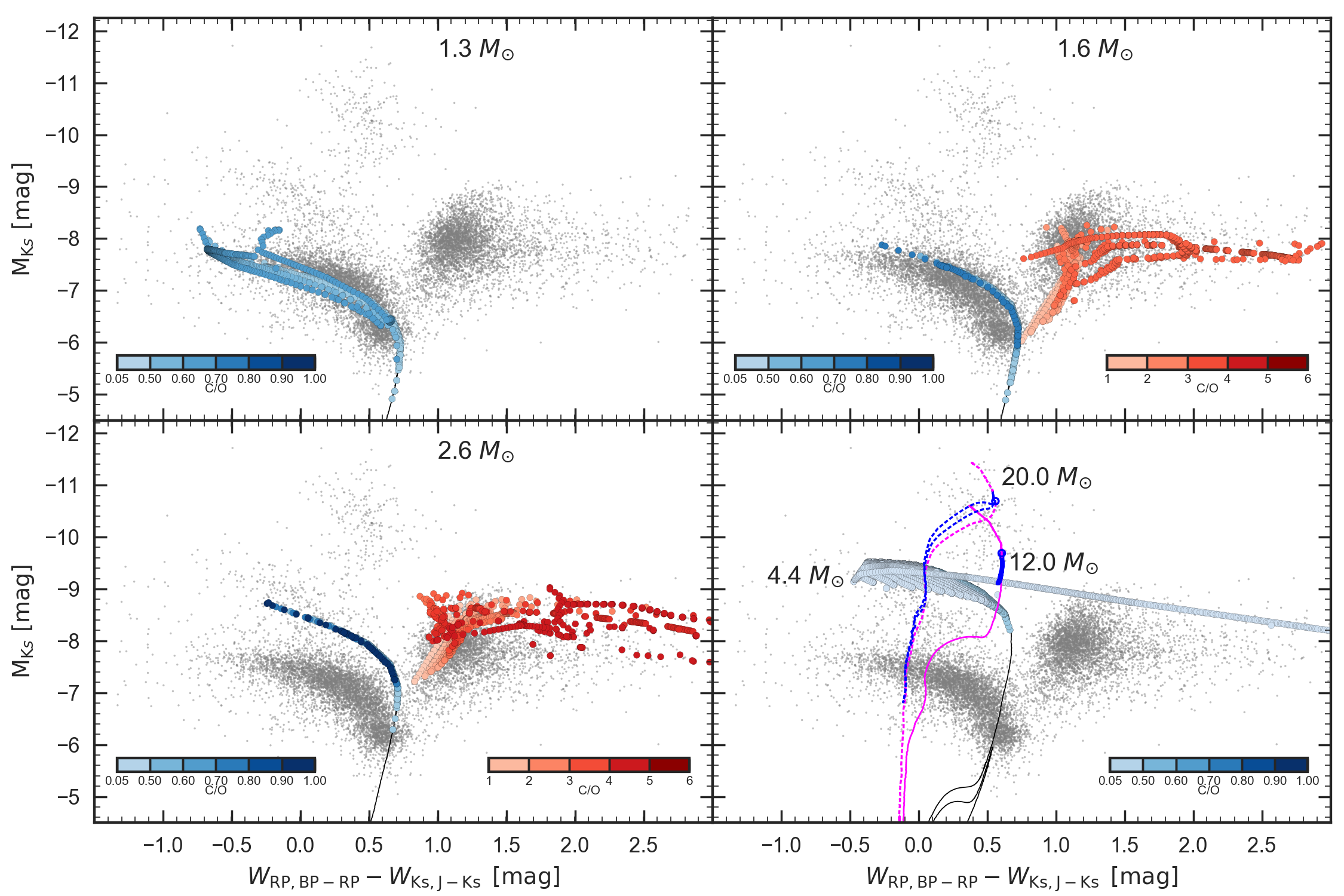}
\caption{Location of stellar evolution tracks computed with the \texttt{PARSEC} and \texttt{COLIBRI} codes \citep{Bressan_etal_2012, Marigo_etal_2013} for a few selected values of the initial mass (as indicated). The initial metallicity is Z=0.006 for the tracks with $M_{\rm i}=$ 1.3, 1.6, 2.6, and 4.4 $M_{\odot}$, and Z=0.008 for the massive ones with $M_{\rm i}=$ \mass{12} (solid line) and \mass{20} (dashed line).
The TP-AGB phase is colour-coded according to the surface C/O ratio, while the previous evolution is shown with a black line. 
In the right-bottom panel the visible sections of the massive star tracks include part of the H-shell burning phase, the core-helium burning phase in blue (a blue loop is visible for the \mass{20} models), and the stages just preceding the ignition of carbon in the core. The circles mark the beginning of the core-helium burning phases in the two massive stars.}
\label{Fig:LMC_WRPminusWK_versus_K_withTracks}
\end{figure*}

During their evolution on the AGB, stars of low and intermediate mass exhibit strong variations of their stellar parameters including temperature, composition and luminosity on various timescales. 
These variations have been extensively studied with the help of stellar evolution models \citep[e.g.][]{1993ApJ...413..641V,Marigo_etal_2008} because of their significant impact on stellar and galactic evolution \citep[e.g.][]{2014PASA...31...30K,Marigo_etal_2017}.
In this appendix we confront predictions from stellar evolution models with the representation of the AGB population in the LMC provided by the newly introduced $W_{RP}-W_K$ versus $K_s$ diagram.

In Fig.\,\ref{Fig:LMC_WRPminusWK_versus_K_withTracks} we follow the evolutionary path of stars of four different masses, \mass{1.3}, \mass{1.6}, \mass{2.6}, and \mass{4.4}, and a typical LMC metallicity (Z=0.006) on the background of LMC-LPVs identified from \Gaia DR2 data.
The four masses are good representatives of the mass ranges we related the various branches in Figs.\,\ref{Fig:LMC_WRPminusWK_versus_K_withBPmRP} and \ref{Fig:model_LMC_WRP_WJK} to. 
The thermal pulses are clearly visible in all four cases. 
The locations of these tracks nicely confirm the picture
in Sect.\,\ref{Sect:LPVcharacterization}.

The TP-AGB evolution of the \mass{1.3} model well covers the lowest branch on the left (branch \textit{a)} of Fig.~\ref{Fig:model_LMC_WRP_WJK}), and is representative of low-mass stars that do not become carbon stars.

The \mass{1.6} and \mass{2.6} models spend part of their AGB evolution as O-rich, and later become carbon stars as a consequence of a few third dredge-up events.
It is interesting to note that, as long as the surface C/O remains below unity during their TP-AGB phase, the two tracks evolve along the distinct leftward branches of the diagram: the low-mass \mass{1.6} model is found at fainter magnitudes, and the intermediate-mass \mass{2.6} model runs brighter (branches \textit{a)} and \textit{c)} of Fig.~\ref{Fig:model_LMC_WRP_WJK}, respectively).
The reader is referred to Sect.~\ref{Sect:LPVcharacterization} for the interpretation of the observed gap between the two sequences.

The maximum leftward excursion of both tracks (most negative $W_{RP}-W_{K}$ value) corresponds to the stage (typically with C/O$\simeq0.8-0.9$) just before the models become C-rich at the next thermal pulse. When such transition occurs, the newly born C-rich models jump on the rightward branch (sequence \textit{b)} in Fig.~\ref{Fig:model_LMC_WRP_WJK}), where they evolve until the end of the AGB. The final, short-lived phases are characterized by significant mass loss and efficient dust formation, which drive the models to populate the region  
of the so-called extreme C-stars (with large $J-K_s$ colours).

The \mass{4.4} model represents the case of a star that experiences HBB. 
While the brightest portion of the Early-AGB takes place at almost constant $W_{RP}-W_{K}$, the subsequent TP-AGB evolution develops on the brighter leftward branch (sequence \textit{c)} in Fig.~\ref{Fig:model_LMC_WRP_WJK}). 
We also expect that HBB stars populate the region spread around the base of the almost vertical branch where the RSG stars are found (branch \textit{d)} in Fig.~\ref{Fig:model_LMC_WRP_WJK}). 
The last steps in the evolution on the AGB, when the star moves to high values in $J-K_s$ due to heavy mass loss, occur on very short time scales for these objects.

To complete the picture of the various mass domains, we plot the evolutionary tracks of a \mass{12} and \mass{20} models with initial metallicity Z=0.008 (bottom right panel of Fig.~\ref{Fig:LMC_WRPminusWK_versus_K_withTracks}).
The two tracks populate the lower and the upper part of branch \textit{d)} (see Fig.~\ref{Fig:model_LMC_WRP_WJK}).

\subsection{Synthetic stellar populations}

We computed a detailed simulation for the stellar populations of the LMC with the aid of the TRILEGAL code \citep{Girardi_etal_2005}. 
To this aim, we adopted the results from \citet{Harris_Zaritsky_2009} to describe the spatially resolved star formation history, age-metallicity relation, distance and reddening for about 400 regions across the LMC. We adopt a Kroupa IMF \citep{Kroupa_2001} and we calibrated the total mass in stars of each sub-field to reproduce the RGB star counts. 
The total mass in stars of the simulation is $\approx2.3\cdot10^9 M_{\sun}$. 
Stellar isochrones, based on extended grids of PARSEC and COLIBRI evolutionary tracks \citep{Bressan_etal_2012, Marigo_etal_2013}, are described in \cite{Marigo_etal_2017}. 
They include the effect on simulated photometry due to circumstellar dust around mass-losing AGB stars. 
Evolution along the Thermally Pulsing AGB is followed up to the complete ejection of the envelope by stellar winds and accounts for the formation of carbon stars due to the third dredge-up, as well as the occurrence of HBB in the most massive TP-AGB stars (with initial masses > 4\,$M_{\sun}$ at the relevant metallicities). 
In this work, we used a set of available TP-AGB tracks calibrated to reproduce star counts and luminosity functions for AGB stars in the SMC (Pastorelli et al., in prep.). 

The simulation is complete down to \gmag = 20 mag. 
In \Gaia DR2, a total number of 11\,022 LPVs in the LMC have magnitudes brighter than $K_s$ = 13.7 mag and \grp = 17.5.
Adopting these limits, our LMC simulation predicts a much higher number of stars, i.e. about 1.69$\cdot 10^5$. 

The excess of simulated stars corresponds to red giants
with variability amplitudes too low for them
to be detected as variables. 
These need to be
excluded from the simulation for a consistent comparison
with observations. In principle, this would require to estimate
variability amplitudes for the stars in the simulation.
However, a simple relation between pulsation
amplitudes and global stellar parameters is
currently unavailable, so that direct computations
with non-linear pulsation models would be necessary.
The main obstacles in the use of non-linear models
are that: (1) to reproduce observations, they
include free parameters, which lack a proper calibration
as a function of global stellar parameters
\citep[see, \textit{e.g.}][and references therein]{Olivier_Wood_2005}, and
(2) they are generally time-consuming,
and thus they cannot be readily implemented in present synthetic populations.

For these reasons, we follow a different approach,
based on the results of linear pulsation models.
To begin with, we exploit the fact that
the dominant mode of pulsation in LPVs
shifts towards lower radial orders as they evolve,
with the fundamental mode being the last
one to become dominant, right after the first
overtone mode. All stars in our observed
sample are pulsating either in the first
overtone or in the fundamental mode (Lebzelter et al., in prep.),
\textit{i.e.}, all of them have evolved to the
point at which the first overtone mode has
become dominant, or past it. To determine
which simulated stars have gone through that point,
we make use of the fact that,
for overtone modes, growth rates are a
good proxy for variability amplitudes \citep{2017ApJ...847..139T}.
To a first approximation, they depend
only on the ratio between the dynamical frequency
and the acoustic cut-off frequency of the star,
proportional to $\tilde{\nu}=(RT_{\rm eff}/M)^{1/2}$ (Trabucchi et al., in prep.).
The dependence is monotonic, and follows a functional
form that is similar in the range of
parameters corresponding to typical AGB stars,
for which one finds $4\lesssim\tilde{\nu}\lesssim18$.
This allows to formulate a selection criterion
in terms of global stellar parameters:
we keep only simulated stars for which
\begin{equation}
    \tilde{\nu}=(RT_{\rm eff}/M)^{1/2} \gtrsim \tilde{\nu}_{\rm min} \simeq 10 \,,
\end{equation}
where the threshold on the right-hand side is estimated
from linear pulsation models. The exact value
depends on the relation between amplitudes and
growth rates, that has not been derived yet.
However, this does not affect our results.
We find that the morphology of the
observed $(W_{RP}-W_K)$ versus $K_s$ diagram
is reasonably well reproduced assuming
$\tilde{\nu}_{\rm min}\simeq8-10$, depending
slightly on mass and chemical type.
Being based on pulsation models appropriate
for AGB stars, the above considerations
cannot be applied to RSGs on branch (d),
that are experiencing the core He-burning phase.
Therefore, no selection is applied to those
stars.
\end{document}